\documentclass[runningheads]{llncs}
\usepackage{eccv}
\usepackage{eccvabbrv}
\usepackage{graphicx}
\usepackage{booktabs}
\usepackage{tabularx}
\usepackage{multirow}
\usepackage[table]{xcolor}
\usepackage{microtype}
\usepackage[accsupp]{axessibility} 
\usepackage{enumitem}
\usepackage{makecell}
\usepackage{hyperref}
\usepackage{orcidlink}

\begin{document}
\title{Benchmarking the Alignment of Data-Quality Metrics, Human Judgment and Land-Cover Segmentation Performance for Earth Observation} 

\titlerunning{Benchmarking the Alignment of Data-Quality Metrics}

\author{Ümit Mert Çağlar\orcidlink{0000-0002-0391-3847} \and
Alptekin Temizel\orcidlink{0000-0001-6082-2573}}
\authorrunning{Ü.M.~Çağlar A.~Temizel}
\institute{Graduate School of Informatics, Middle East Technical University, Ankara, Turkey
\email{\{mert.caglar,atemizel\}@metu.edu.tr}}

\maketitle

\begin{abstract}
Volume and quality of datasets are crucial for deep learning model training, yet they are often constrained by availability and data acquisition costs. Synthetic data augmentation can extend existing datasets with realistic images, and the quality of these images is generally assessed  through fidelity metrics such as FID, KID, IS, LPIPS and SSIM that measure structural or distributional similarity. However, such metrics, including the widely used FID, focus on visual fidelity without reflecting downstream utility, and can diverge from human perception under perturbations that are imperceptible to human observers. In this work, we systematically evaluate Earth observation datasets alongside synthetic counterparts generated by deep generative models, comparing automatic metrics against human perception and downstream tasks. Our results reveal a stark misalignment: semantics-preserving perturbations such as rotation drastically alter metric scores while leaving human recognition unaffected, and synthetic samples that score poorly on automatic metrics achieve comparable or higher perceived realism, and can improve downstream performance when combined with real data. By benchmarking semantic segmentation models trained on mixed real-synthetic datasets, we demonstrate that quality metrics rooted in ImageNet-pretrained feature spaces are unreliable indicators for geospatial data. Our findings underscore that automatic quality evaluation of synthetic datasets should be grounded in downstream task performance and human evaluation.
\end{abstract}

\section{Introduction}
\label{sec:intro}

Deep generative frameworks, such as Denoising Diffusion Probabilistic Models (DDPMs) and Generative Adversarial Networks (GANs), have enabled synthetic data augmentation across various computer vision tasks. But assessing their actual utility remains challenging. Standard evaluation metrics focus on image fidelity and diversity, which are often decoupled from downstream task performance. Consequently, effectively measuring synthetic data quality in terms of downstream application utility remains an open problem.

Standard evaluation practice relies on distribution-level distances between synthetic and real datasets, computed in the latent space of a pre-trained network. The Fréchet Inception Distance (FID)~\cite{heusel2017gans} in particular has become a standard benchmark for generative model evaluation. Although FID performs well in many natural image settings and often aligns with human perception, it exhibits several well-documented failure modes. As a result, several complementary metrics are typically reported together. Furthermore, latent-distance metrics rely on feature extractors of models pre-trained on generic datasets, most commonly ImageNet~\cite{deng2009imagenet}, whose learned representations may not transfer to specialized domains. 

Remote sensing data for Earth observation is one such domain where assumptions and feature representations learned from ImageNet, shaped by object-centric composition and natural viewing geometry, do not necessarily hold true. Unlike natural images, Earth observation imagery is acquired from a top-down viewpoint, has arbitrary orientation, and  covers large geographical regions in which semantic information is distributed across the scene. As a consequence, the validity of commonly used quality metrics becomes questionable as they implicitly assume that proximity to a reference dataset is indicative of downstream utility. However, in practice, synthetic data may achieve favorable scores through structural memorization or low-frequency smoothing while suppressing high-frequency details that downstream tasks rely upon~\cite{Adamkiewicz_2026_CVPR}. Conversely, transformations such as rotations and flips can substantially alter metric values despite leaving the underlying semantic structure unchanged.

To analyze the alignment between quantitative evaluation metrics, human perception and downstream task utility, we combined perturbation experiments, human perception study and data quality metric benchmarks. Perturbation experiments measure the response of seven quantitative metrics to different image transformations. Human perception study covers a four-stage questionnaire comprising a perturbation evaluation, image-segmentation pair matching, conditionally generated content preference and realism assessment of different synthetic model outputs. These experiments are conducted on real and synthetic remote sensing imagery, containing Earth observation for land-cover segmentation.

\noindent Our primary contributions are summarized as follows:
\begin{itemize}[topsep=3pt, itemsep=3pt, parsep=0pt, leftmargin=*]
    \item \textbf{Vulnerability of generative metrics:} We show that widely
    used latent-space distribution metrics can be substantially altered by
    minor or semantics-preserving image perturbations, motivating more robust
    evaluation paradigms.
    \item \textbf{Human perception study:} We conduct a four-stage study with
    88 participants quantifying recognition accuracy, visual preference, and
    cognitive load across real and synthetic Earth observation datasets,
    providing a human reference for evaluating quality metrics.
    \item \textbf{Three-way alignment analysis:} We quantify the correlations
    and discrepancies among human perception, quality metrics, and
    downstream model performance, exposing failure modes of standard
    automated evaluation.
    \item \textbf{Land-cover segmentation benchmark:} We benchmark model
    performance when trained on varying mixtures of real and synthetic data,
    revealing that visual fidelity does not reliably predict downstream
    performance and that synthetic data can improve real-data baselines despite
    poor fidelity scores.
\end{itemize}

\section{Related Work}

Early generative model evaluation relied on qualitative inspection and likelihood-based estimates, both of which scale poorly~\cite{theis2015note}. While pair-wise metrics like the Structural Similarity Index Measure (SSIM)~\cite{wang2004image} evaluate low-level features (luminance, contrast, and structure), modern evaluation favors automated, distribution-based measures. The Inception Score~\cite{salimans2016improved} pioneered this shift, leading to Fréchet Inception Distance (FID)~\cite{heusel2017gans}, which computes the distance between real and synthetic feature representations. To eliminate FID's sample size bias, the Kernel Inception Distance (KID)~\cite{binkowski2018demystifying} applies the squared Maximum Mean Discrepancy (MMD) to these distributions. Alternatively, the Learned Perceptual Image Patch Similarity (LPIPS)~\cite{zhang2018unreasonable} metric assesses fine-grained perceptual similarity by comparing normalized deep features across neural network layers.

Despite their popularity, automated synthetic quality metrics have several limitations due to their dependence on pre-trained ImageNet feature extractors. Because these representations are optimized for ImageNet domain, they may introduce class-dependent biases and transfer poorly to different domains~\cite{kynkaanniemi2022role}. This issue is particularly pronounced in specialized settings such as facial imagery, where feature embeddings may overemphasize certain visual attributes while failing to fully capture perceptual quality and diversity~\cite{cetin2024facial}.

Beyond representational bias, FID is also sensitive to implementation details and low-level perturbations. Variations in preprocessing, such as resizing or interpolation choices, can significantly affect scores, raising concerns about reproducibility~\cite{parmar2022aliased}. Moreover, because FID operates in a high-dimensional feature space, imperceptible image perturbations may lead to large metric variations, while meaningful structural distortions may remain undetected~\cite{jia2026evaluating}. Standardization efforts such as Clean-FID~\cite{parmar2022aliased} address these inconsistencies by enforcing consistent preprocessing pipelines.

The limitations of purely distribution-based metrics have become particularly apparent in text-to-image generation, where image quality alone is insufficient and alignment with the conditioning text is required. To address this, CLIP-based evaluation methods have been proposed. CLIP-Score measures the cosine similarity between text and image embeddings in a shared representation space learned from large-scale image-text data~\cite{hessel2021clipscore}. Similarly, CLIP-FID replaces the Inception backbone in FID with CLIP encoders, leading to improved correlation with human judgments in diverse generative settings~\cite{jayasumana2024rethinking}.

Beyond perceptual quality, recent studies highlighted the importance of evaluating semantic reliability and task utility. In text-to-image generation, uncertainty estimation frameworks have been introduced to quantify confidence and semantic consistency of generated outputs~\cite{franchi2025towards}. In privacy-preserving synthetic data generation, particularly in medical domain, there is an inherent trade-off between fidelity and privacy. Models without differential privacy (DP) can achieve high visual realism and downstream utility but may introduce re-identification risks~\cite{adams2025fidelity}. Conversely, enforcing DP disrupts statistical structure, often leading to substantial degradation in downstream predictive performance~\cite{miletic2025utility}. These findings suggest that high perceptual fidelity, measured by low FID scores, does not necessarily imply high task utility.

Recent literature heavily explores generative models to mitigate data scarcity and high annotation costs in Earth observation (EO) systems~\cite{huang2026survey}. To introduce greater visual and semantic diversity beyond traditional transformations, several frameworks~\cite{sousa2025data} leverage diffusion models to combine meta-prompts and vision-language models to fine-tune EO-specific diffusion pipelines. Similarly, EarthSynth~\cite{pan2025earthsynthgeneratinginformativeearth} is a generative foundation model trained on a multi-task dataset to synthesize multi-category, cross-satellite labeled imagery for downstream interpretation. Other approaches focus on targeted spatial augmentation, such as automatically generating and blending specific objects into existing satellite scenes to boost object detection performance~\cite{khammari2024synthetic}. Furthermore, synthetic data estimates are also applied to multimodal sensor-fusion datasets, leveraging generative modeling to supplement sparse temporal and spatial coverage in remote regions~\cite{mutakabbir2026noah}. Different synthetic modes, specifically physics engine renders and generative models are shown to provide significant performance gains for downstream tasks~\cite{hisam2025impact}.

Despite these advancements, existing evaluation methodologies for generative models remain susceptible to biases, perturbations and limitations. Although distribution-based quality metrics like FID and its variants can evaluate visual fidelity, task-specific downstream utility or conceptual alignment assessment are not available. Fidelity, utility and perception alignment are rarely unified within a single framework, limiting quality metrics for domain-specific settings such as Earth observation. Crucially, the literature lacks a systematic investigation that analyzes how automatic quality metrics, human perception and downstream task performance align, and more importantly, a conceptual automatic metric that evaluates utility and fidelity together.

To address this gap, we evaluate three complementary sources of evidence for image quality on a common collection of real and synthetic Earth observation datasets: (i)~automatic quality metrics under controlled image perturbations, (ii)~human perception through a four-stage user study, and (iii)~downstream land-cover segmentation performance. A joint analysis of these sources of evidence on the same datasets reveals where they align and, more importantly, where they diverge. We first describe the datasets and perturbations used throughout the study.

\section{Datasets and Perturbations}
\label{sec:samples}
The set of real and synthetic remote sensing datasets used in our evaluation are summarized in Table~\ref{tab:datasets}. ARAS400k~\cite{ccauglar2026grounding} is a high-resolution remote sensing dataset covering diverse urban and rural landscapes across Türkiye at 10-meter spatial resolution, with corresponding land-cover segmentation maps. BELDE (Building a Large-scale Earth-observation Land-cover Dataset for European Subcontinent)~\cite{caglar2026belde} contains over one million image-mask pairs. We further evaluate on BELDE-K (Republic of Korea) and BELDE-CA-NV (California-Nevada), which follow the same annotation format but represent geographically distinct regions with different land-cover distributions, enabling evaluation under domain shift. We use the same training and test splits.

Synthetic samples, shown in Figure~\ref{tab:image_comparison}, are generated using two segmentation-conditioned architectures and one unconditional model, each trained on the corresponding real dataset. These include ARAS-CSD and BELDE-CSD (segmentation conditioned Stable Diffusion), ARAS-CUGAN (conditional U-Net GAN), and ARAS-SGAN3 (unconditional StyleGAN3) and ARAS-SGAN3-D (diverse coreset of ARAS-SGAN3).

\begin{table}[t]
  \centering
  \caption{Datasets used in this work. Synthetic datasets are shaded. All patches are $256\times256$ pixels at 10-meter spatial resolution and follow a unified preprocessing pipeline with seven shared land-cover classes.}
  \label{tab:datasets}
\begin{tabular}{llllr}
    \toprule
    Name & Type & Generator & Region & \#Patches \\
    \midrule
    ARAS-train    & Real          & ---              & Türkiye           & 80,192    \\
    ARAS-test     & Real          & ---              & Türkiye           & 10,024    \\
    BELDE         & Real          & ---              & Europe            & 1,088,385 \\
    BELDE-K       & Real          & ---              & Rep.\ of Korea    & 16,607    \\
    BELDE-CA-NV   & Real          & ---              & California-Nevada & 88,155    \\
    \midrule
    \rowcolor{gray!12}
    ARAS-CSD      & Conditional   & Stable Diffusion & Türkiye           & 80,192    \\
    \rowcolor{gray!12}
    BELDE-CSD     & Conditional   & Stable Diffusion~~ & Türkiye            & 80,192    \\
    \rowcolor{gray!12}
    ARAS-CUGAN    & Conditional   & UNet-GAN         & Türkiye           & 80,192    \\
    \rowcolor{gray!12}
    ARAS-SGAN3    & Unconditional~~ & StyleGAN3        & Türkiye           & 300,000   \\
    \rowcolor{gray!12}
    ARAS-SGAN3-D~~  & Unconditional & StyleGAN3        & Türkiye           & 100,000   \\
    \bottomrule
  \end{tabular}
\end{table}

All image patches used in this work undergo identical pre-processing and normalization steps. The associated land-cover segmentation maps consistently use seven distinct classes across all datasets to ensure a standardized benchmarking protocol. However, datasets from different geographic regions exhibit inherently different data distributions. 

\begin{figure}[ht]
\centering
\scriptsize
\begin{tabular}{ccccccc}
\textbf{Real} & \textbf{Noise} & \textbf{Perspective} & \textbf{Rotation} & \textbf{Flip} & \textbf{Resizedcrop} & \textbf{Combined} \\
\includegraphics[width=1.64cm]{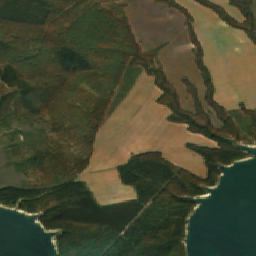} & \includegraphics[width=1.64cm]{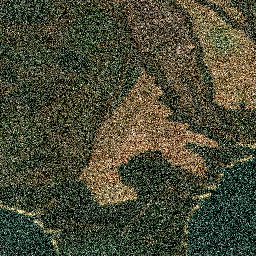} & \includegraphics[width=1.64cm]{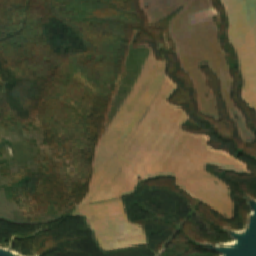} & \includegraphics[width=1.64cm]{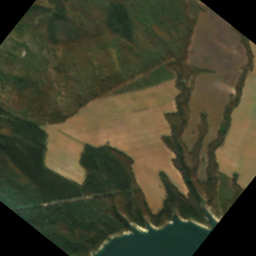} & \includegraphics[width=1.64cm]{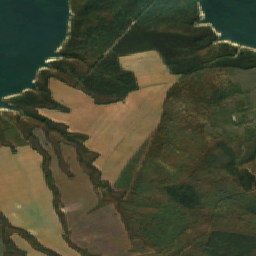} & \includegraphics[width=1.64cm]{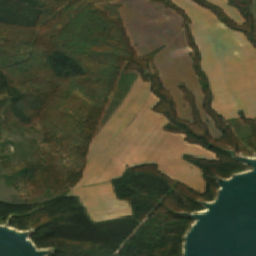} & \includegraphics[width=1.64cm]{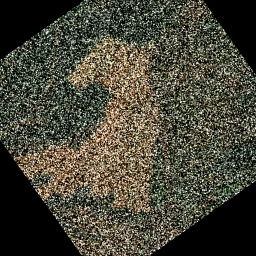} \\
\includegraphics[width=1.64cm]{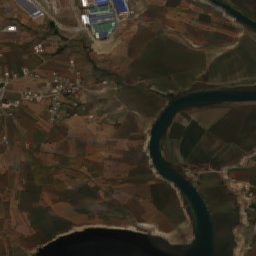} & \includegraphics[width=1.64cm]{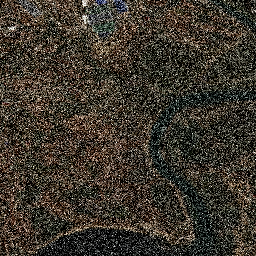} & \includegraphics[width=1.64cm]{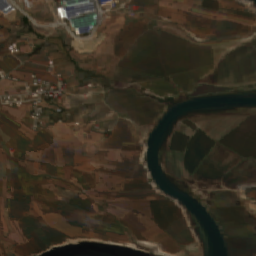} & \includegraphics[width=1.64cm]{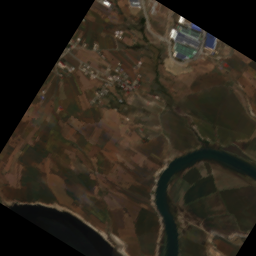} & \includegraphics[width=1.64cm]{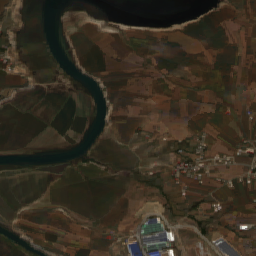} & \includegraphics[width=1.64cm]{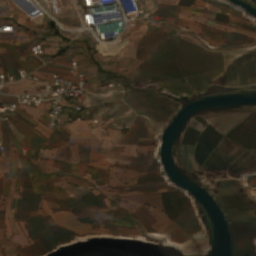} & \includegraphics[width=1.64cm]{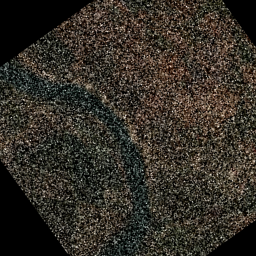} \\
\end{tabular}
\caption{Visual comparison of image transformations across real images.}
\label{tab:visual_comparison}
\end{figure}

As visualized in Figure~\ref{tab:visual_comparison}, we apply six transform operations to systematically
characterize metric variations with respect to human detection limits.
\textbf{Noise} adds Gaussian noise with dynamically scaled variance,
simulating sensor artifacts and atmospheric interference.
\textbf{Perspective} applies random projective warps modeling off-nadir
viewpoints, with scale distortion in $[0.05, 0.1]$.
\textbf{Rotation} applies random affine transformations with scaling in
$[0.9, 1.1]$, $\pm 10\%$ translation, and rotation
$\theta \in [-45^\circ, 45^\circ]$ under bilinear interpolation.
\textbf{Flip} performs horizontal and vertical reflections, preserving
semantic content.
\textbf{ResizedCrop} randomly crops $50$--$100\%$ of the image area with
aspect ratios in $[0.75, 1.33]$, then bilinearly resizes to
$256 \times 256$ pixels.
\textbf{Combined} sequentially composes cropping, flipping, perspective
distortion, noise, and affine transformation, inducing severe compounded
structural and spectral degradation.

\section{Human Perception Study}

\begin{figure}[htbp]
    \centering

    \begin{subfigure}[b]{0.48\textwidth}
        \centering
        \includegraphics[width=\linewidth]{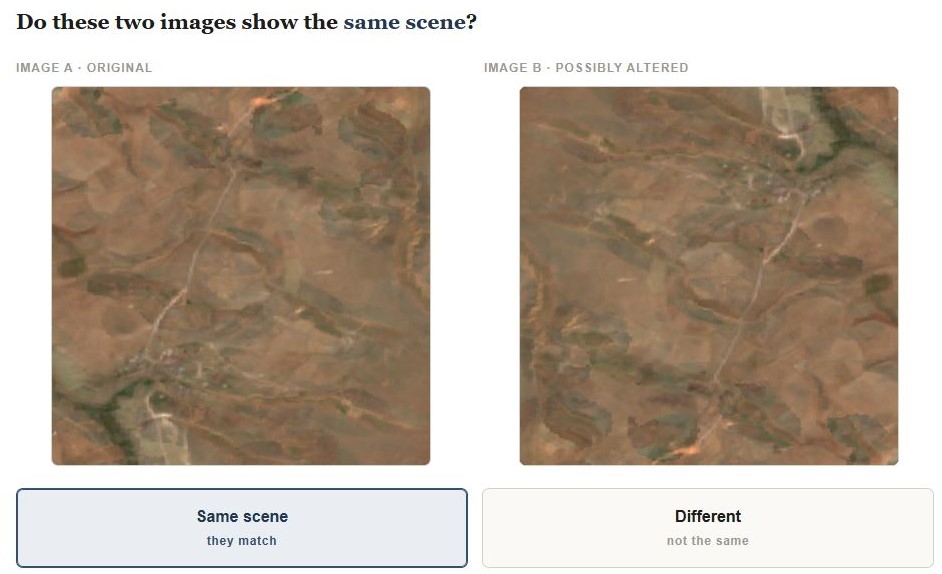}
        \caption{Augmentation alignment}
        \label{fig:a}
    \end{subfigure}
    \hfill
    \begin{subfigure}[b]{0.48\textwidth}
        \centering
        \includegraphics[width=\linewidth]{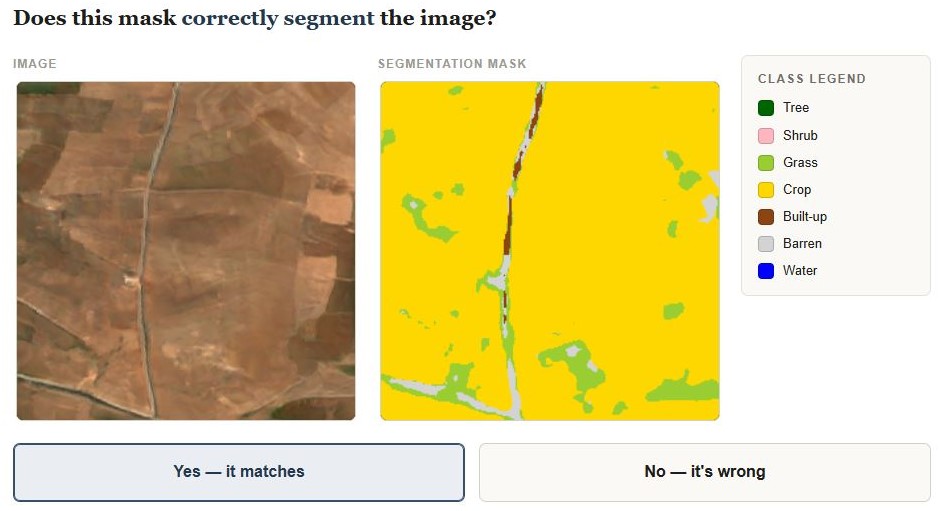}
        \caption{Utility for downstream tasks}
        \label{fig:b}
    \end{subfigure}

    \vspace{0.5em}

    \begin{subfigure}[b]{0.48\textwidth}
        \centering
        \includegraphics[width=\linewidth]{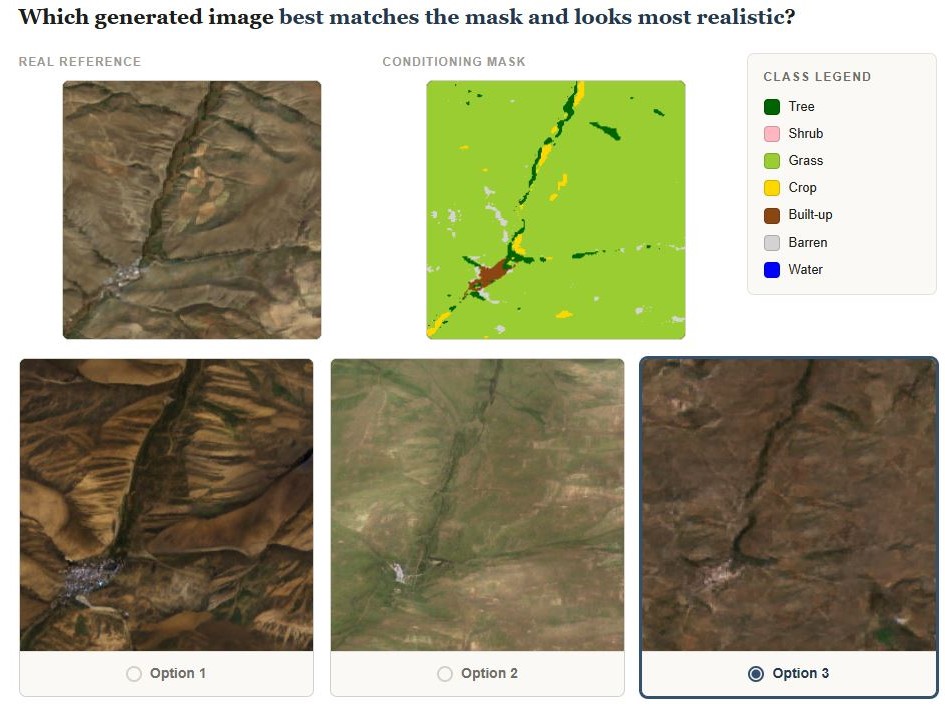}
        \caption{Conditional generation preference}
        \label{fig:c}
    \end{subfigure}
    \hfill
    \begin{subfigure}[b]{0.48\textwidth}
        \centering
        \includegraphics[width=\linewidth]{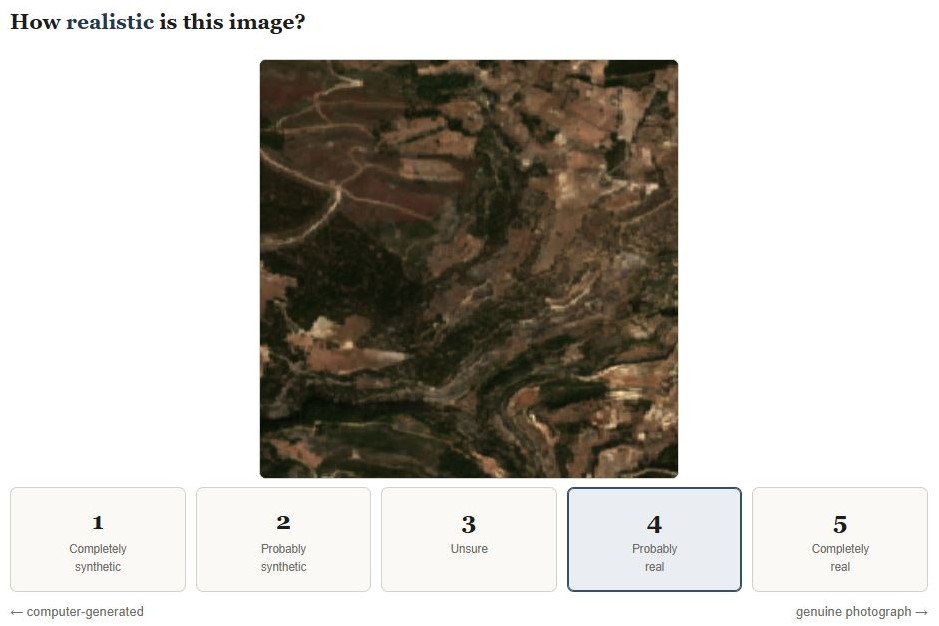}
        \caption{Data realism scores}
        \label{fig:d}
    \end{subfigure}

    \caption{Human perception study screens for (a) data augmentation alignment, (b) utility for downstream tasks, (c) conditional generation preference and (d) data realism scores}
    \label{fig:main}
\end{figure}

The perception study evaluates the alignment between human perception, automated quality metrics, and downstream data utility using samples from the datasets outlined in Table~\ref{tab:datasets}. As illustrated in the comprehensive pipeline (Fig.~\ref{fig:main}), the experiment is structured into four sequential phases designed to quantify distinct dimensions of visual and semantic fidelity, while concurrently recording participant choices and reaction times to assess cognitive load.

The initial phase establishes a baseline for human perception by evaluating data augmentation alignment (Fig.~\ref{fig:a}), asking participants to determine whether a baseline image and its altered variant depict the identical scene. The specific augmentations evaluated are detailed in Figure~\ref{tab:visual_comparison}, and the task incorporates two distinct control groups: unaltered pairs of the same scene and pairs drawn from entirely different scenes.

Shifting from geometric alterations to semantic consistency, the second phase measures data utility for downstream tasks (Fig.~\ref{fig:b}) by tasking participants with judging the alignment between an Earth observation image and its associated land-cover segmentation map. To test whether real and synthetic samples yield equal interpretability, these image-mask pairs are split evenly between real data and their synthetic counterparts. Furthermore, deliberately mismatched image-mask pairs are introduced as negative trials; this balanced inclusion of valid and invalid pairs ensures that participant accuracy reflects genuine semantic discrimination rather than an acquiescence bias.

\begin{figure}[htbp]
\centering
\scriptsize
\begin{tabular}{ccccc}
    \textbf{Real Images} & \textbf{Conditioning} & \textbf{ARAS-CDS} & \textbf{BELDE-CDS} &\textbf{ARAS-CUGAN}\\
    \includegraphics[width=2.3cm]{samples/sreals/0579.png} & \includegraphics[width=2.3cm]{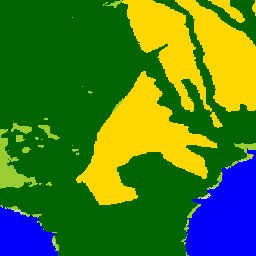} & \includegraphics[width=2.3cm]{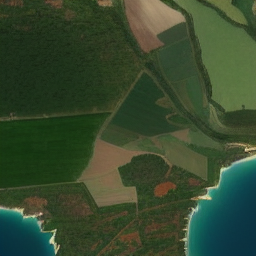} & \includegraphics[width=2.3cm]{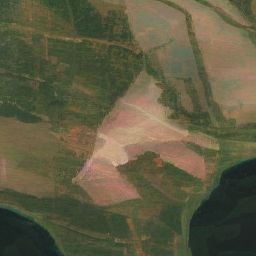} & \includegraphics[width=2.3cm]{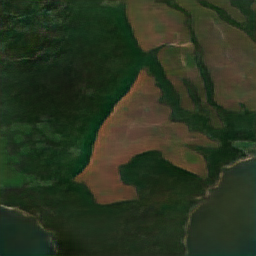} \\
    \includegraphics[width=2.3cm]{samples/sreals/0722.png} & \includegraphics[width=2.3cm]{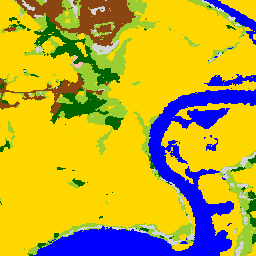} & \includegraphics[width=2.3cm]{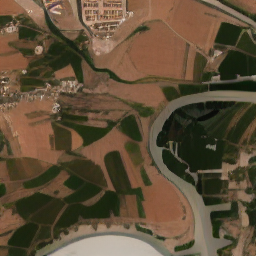} & \includegraphics[width=2.3cm]{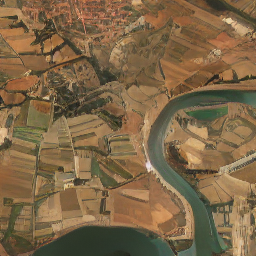} & \includegraphics[width=2.3cm]{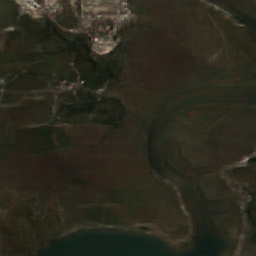} \\
    \includegraphics[width=2.3cm]{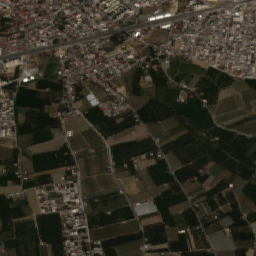} & \includegraphics[width=2.3cm]{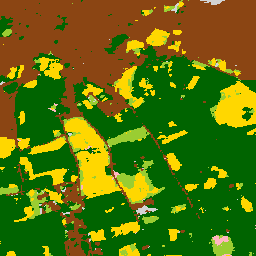} & \includegraphics[width=2.3cm]{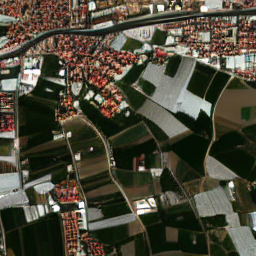} & \includegraphics[width=2.3cm]{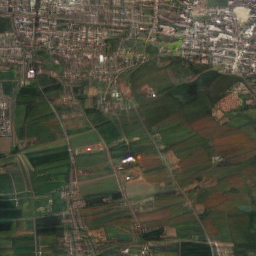} & \includegraphics[width=2.3cm]{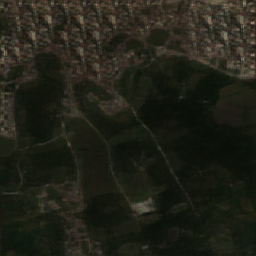} \\
\end{tabular}
\caption{Image generation comparison of conditional Stable Diffusion for BELDE-trained (BELDE-CSD), ARAS-trained (ARAS-CSD) and conditional U-Net GAN (ARAS-CUGAN) with the real and conditioning images (segmentation)}
\label{tab:image_comparison}
\end{figure}

The third phase investigates conditional generation preferences (Fig.~\ref{fig:c}) using the generative samples illustrated in Figure~\ref{tab:image_comparison}. Given a real Earth observation image and its corresponding ground-truth segmentation map as conditioning inputs, participants assess the perceived quality and realism of the generated outputs. This phase specifically aims to isolate which conditional generation architectures produce the most consistent and contextually plausible imagery according to human evaluators.

Finally, the study concludes with the collection of explicit data realism scores (Fig.~\ref{fig:d}). Participants grade individual images, sampled across both real and synthetic datasets, on a 1-to-5 Likert scale. This subjective valuation is designed to determine the correlation between quantitative automated metrics and the qualitative, human assessment of Earth observation datasets.

\section{Results}
The evaluation commenced with an initial cohort of 95 unique participants and 5769 images. Quality control filtering dropped 7 participants due to failed attention checks. To handle localized anomalies without losing entire trials, 134 instances with extreme reaction time outliers, according to the $3 \times \text{IQR}$ rule, were imputed with the mean reaction time.

\begin{table}[htbp]
\centering
\caption{Perceived scene understanding across different augmentations.}
\label{tab:image_transforms}
\small 
\setlength{\tabcolsep}{10pt} 
\begin{tabular}{lccc}
\toprule
\textbf{Transform} & 
\textbf{\makecell[c]{Accuracy\\(\%)}} & 
\textbf{\makecell[c]{Avg. Reaction\\Time (s)}} & 
\textbf{\makecell[c]{Total\\Trials}} \\
\midrule
Baseline     & 97.13 & 6.35 & 174 \\
Rotation     & 94.05 & 6.42 & 252 \\
Perspective  & 92.22 & 6.94 & 257 \\
Flip         & 91.80 & 6.57 & 256 \\
Resized Crop & 90.44 & 6.51 & 272 \\
Noise        & 79.20 & 6.04 & 250 \\
Combined     & 56.63 & 8.10 & 279 \\
\bottomrule
\end{tabular}
\end{table}

Table~\ref{tab:image_transforms} shows the results of the first stage, highlighting that noise and combined perturbations are the most harmful to human scene perception, while rotation, perspective, flip, and resized crop are the least. The unperturbed baseline yields the highest accuracy (97.13\%), and the heaviest perturbation, the combined transform, reduces accuracy to 56.63\%. Notably, geometric transforms leave human accuracy nearly intact (e.g., 94.05\% under rotation). 

The results of the second stage show that humans verify real and synthetic image-mask pairs with comparable accuracy (72.87\% vs.\ 72.41\%) and similar reaction times (9.17 s vs. 9.28 s), indicating synthetic land-cover masks are as interpretable to human evaluators as real ones. 

Third stage results for perceived image quality and consistency for three conditional generative models indicate conditional Stable Diffusion trained on ARAS400k dataset (ARAS-CSD) is preferred the most often (39\%) with BELDE-CSD slightly behind at 33\%. The lowest preference is for the outputs of ARAS-C-UNetGAN with 28\%, still competitive with other models.

\begin{table}[htbp]
\centering
\caption{Realism scores in Likert-scale for different datasets}
\label{tab:realism_likert}
\small 
\setlength{\tabcolsep}{8pt} 
\begin{tabular}{llcccc}
\toprule
\textbf{Dataset} & \textbf{Type} & \textbf{$\mu$} & \textbf{Md.} & \textbf{$\sigma$} & \textbf{\makecell[c]{Avg. Reaction\\Time (s)}} \\
\midrule
ARAS            & Real  & 3.54 & 4.0 & 1.13 & 4.97 \\
ARAS-SGAN3      & Synth & 3.43 & 4.0 & 1.11 & 4.52 \\
BELDE-K         & Real  & 3.42 & 4.0 & 1.17 & 5.27 \\
BELDE           & Real  & 3.32 & 4.0 & 1.28 & 4.58 \\
ARAS-SGAN3-D    & Synth & 3.28 & 4.0 & 1.19 & 4.75 \\
BELDE-CA-NV     & Real  & 3.09 & 3.0 & 1.24 & 4.53 \\
ARAS-CUGAN      & Synth & 2.96 & 3.0 & 1.27 & 4.59 \\
BELDE-CSD       & Synth & 2.80 & 3.0 & 1.36 & 4.43 \\
ARAS-CSD        & Synth & 2.51 & 2.0 & 1.32 & 5.44 \\
\bottomrule
\end{tabular}
\end{table}

The final-stage results, summarized in Table~\ref{tab:realism_likert}, present the mean realism scores based on a 5-point Likert scale (1: Most Synthetic, 5: Most Realistic). Results highlight that real datasets exhibit higher realism scores than most of the synthetic datasets. Real datasets generally score higher, but the gap is not clean: the synthetic ARAS-SGAN3 dataset ($\mu = 3.43$) is rated more realistic than two of the four real datasets, BELDE ($\mu = 3.32$) and BELDE-CA-NV ($\mu = 3.09$), despite its substantially worse FID. The Stable Diffusion and U-Net GAN datasets, by contrast, receive the lowest realism ratings. Notably, ARAS-CSD, while being the most preferred model (39\%) when judged against a reference mask, receives the lowest standalone realism score ($2.51\pm1.32$), indicating that fidelity to a conditioning mask and perceived realism of the generated image are distinct, and sometimes opposing, properties.

\subsection{Evaluation Metrics}
Results of automatic referenced distribution-based metrics (FID, KID), non-referenced distribution-based metric (IS) and referenced pair-wise metrics (LPIPS, SSIM and PSNR) across different transform types and target datasets are shown in Table~\ref{tab:metrics_evaluation_reduced_target}. We provide referenced pair-wise metrics for completeness, though we note that pairings are random for some datasets. True pair-wise matching is only applicable to the conditionally generated BELDE-CSD, ARAS-CSD, and ARAS-CUGAN datasets. Baseline results indicate no perturbation, showing clear performance metric for the given target dataset. ARAS-train subset is utilized as the reference dataset for referenced metrics. Flip, resized crop and perspective augmentations exhibit stable quantitative results across the evaluation metrics, albeit small discrepancies. Rotation, noise and combined augmentations are the most detrimental perturbations across the evaluation metrics.

A notable discrepancy is present with the Inception Score (IS) relative to referenced fidelity metrics such as FID. Addition of noise decreases IS slightly, as opposed to the drastic changes in FID scores across the board, conversely IS improves for rotation augmentations, indicating a potential metric-hacking problem. SSIM and PSNR results closely follow the pattern of FID for augmentations, however, they exhibit a stark contrast to FID score when comparing two different datasets. For example, ARAS-CUGAN attains 72 FID, 2.4 IS, 0.46 LPIPS, 0.6 SSIM and 18.71 PSNR when compared against the ARAS-train, meanwhile ARAS-test obtains 2 FID, 3.79 IS, 0.6 LPIPS, 0.38 SSIM and 13.84 PSNR scores. ARAS-CUGAN, a synthetic dataset obtains higher scores in SSIM and PSNR compared to a real dataset. Furthermore, BELDE-CA-NV, a real dataset exhibit 28.68 FID, 4.07 IS, 0.58 LPIPS, 0.38 SSIM and 13.18 PSNR scores, indicating even real datasets can achieve worse FID, IS, LPIPS, SSIM and PSNR scores compared to synthetic ones such as ARAS-SGAN3 or ARAS-CSD. According to our results, none of the automatic metrics provide stable evaluation for real or synthetic datasets.

\begin{table}[htbp]
\centering
\caption{Referenced distribution-based metrics (FID, KID), non-referenced distribution-based metric (IS) and referenced pair-wise metrics (LPIPS, SSIM and PSNR) across different transform types and target datasets. Note: Pair-wise metrics are calculated for matching (synthetic-real) scenes for only ARAS-CUGAN, ARAS-CSD and BELDE-CSD, unconditional random sampling pair-wise calculations are shown as shaded.}
\label{tab:metrics_evaluation_reduced_target}
\scriptsize
\renewcommand{\arraystretch}{0.9}
\begin{tabular}{ll|cc|c|ccc}
\textbf{Target} & \textbf{Transform} & \textbf{FID} $\downarrow$ & \textbf{KID} $\downarrow$ & \textbf{IS} $\uparrow$ & \textbf{LPIPS} $\downarrow$ & \textbf{SSIM} $\uparrow$ & \textbf{PSNR} $\uparrow$ \\
\midrule
& Baseline & 2.09 & 0.00 & 3.79 & \cellcolor{gray!11}0.60 & \cellcolor{gray!11}0.38 & \cellcolor{gray!11}13.84 \\
& Noise & 54.98 & 0.03 & 3.72 & \cellcolor{gray!11}0.66 & \cellcolor{gray!11}0.28 & \cellcolor{gray!11}12.26 \\
& Perspective & 2.84 & 0.00 & 3.75 & \cellcolor{gray!11}0.61 & \cellcolor{gray!11}0.39 & \cellcolor{gray!11}13.85 \\
ARAS-Test & Rotation & 36.51 & 0.02 & 4.75 & \cellcolor{gray!11}0.61 & \cellcolor{gray!11}0.37 & \cellcolor{gray!11}13.60 \\
& Flip & 2.33 & 0.00 & 3.75 & \cellcolor{gray!11}0.60 & \cellcolor{gray!11}0.38 & \cellcolor{gray!11}13.84 \\
& Resized Crop & 2.28 & 0.00 & 3.72 & \cellcolor{gray!11}0.60 & \cellcolor{gray!11}0.39 & \cellcolor{gray!11}13.85 \\
& Combined & 89.86 & 0.05 & 4.74 & \cellcolor{gray!11}0.66 & \cellcolor{gray!11}0.27 & \cellcolor{gray!11}12.22 \\
\midrule
& Baseline & 28.68 & 0.02 & 4.07 & \cellcolor{gray!11}0.58 & \cellcolor{gray!11}0.38 & \cellcolor{gray!11}13.19 \\
& Noise & 82.79 & 0.05 & 3.90 & \cellcolor{gray!11}0.65 & \cellcolor{gray!11}0.27 & \cellcolor{gray!11}11.63 \\
& Perspective & 30.47 & 0.02 & 4.02 & \cellcolor{gray!11}0.58 & \cellcolor{gray!11}0.39 & \cellcolor{gray!11}13.20 \\
BELDE-CA-NV & Rotation & 59.92 & 0.03 & 5.11 & \cellcolor{gray!11}0.59 & \cellcolor{gray!11}0.37 & \cellcolor{gray!11}12.95 \\
& Flip & 29.34 & 0.02 & 4.06 & \cellcolor{gray!11}0.58 & \cellcolor{gray!11}0.38 & \cellcolor{gray!11}13.19 \\
& Resized Crop & 29.29 & 0.02 & 4.02 & \cellcolor{gray!11}0.58 & \cellcolor{gray!11}0.39 & \cellcolor{gray!11}13.20 \\
& Combined & 109.40 & 0.06 & 4.99 & \cellcolor{gray!11}0.65 & \cellcolor{gray!11}0.27 & \cellcolor{gray!11}11.68 \\
\midrule
& Baseline & 54.14 & 0.04 & 4.06 & \cellcolor{gray!11}0.62 & \cellcolor{gray!11}0.39 & \cellcolor{gray!11}13.67 \\
& Noise & 93.59 & 0.06 & 3.99 & \cellcolor{gray!11}0.67 & \cellcolor{gray!11}0.28 & \cellcolor{gray!11}12.24 \\
& Perspective & 53.86 & 0.04 & 3.97 & \cellcolor{gray!11}0.62 & \cellcolor{gray!11}0.39 & \cellcolor{gray!11}13.68 \\
BELDE-K & Rotation & 83.80 & 0.06 & 4.85 & \cellcolor{gray!11}0.63 & \cellcolor{gray!11}0.37 & \cellcolor{gray!11}13.43 \\
& Flip & 53.62 & 0.04 & 4.06 & \cellcolor{gray!11}0.62 & \cellcolor{gray!11}0.39 & \cellcolor{gray!11}13.67 \\
& Resized Crop & 53.07 & 0.04 & 4.03 & \cellcolor{gray!11}0.62 & \cellcolor{gray!11}0.39 & \cellcolor{gray!11}13.67 \\
& Combined & 120.27 & 0.07 & 4.78 & \cellcolor{gray!11}0.68 & \cellcolor{gray!11}0.27 & \cellcolor{gray!11}12.21 \\
\midrule
\midrule

& Baseline & 26.23 & 0.02 & 4.06 & 0.59 & 0.40 & 14.31 \\
& Noise & 77.00 & 0.04 & 3.90 & 0.65 & 0.28 & 12.32 \\
& Perspective & 24.90 & 0.02 & 4.02 & 0.59 & 0.41 & 14.32 \\
BELDE-CSD & Rotation & 52.90 & 0.03 & 5.01 & 0.59 & 0.38 & 13.99 \\
& Flip & 26.34 & 0.02 & 4.03 & 0.59 & 0.40 & 14.31 \\
& Resized Crop & 26.05 & 0.02 & 4.06 & 0.59 & 0.40 & 14.31 \\
& Combined & 99.13 & 0.05 & 4.89 & 0.65 & 0.28 & 12.35 \\
\midrule
& Baseline & 51.42 & 0.04 & 6.19 & 0.63 & 0.35 & 13.09 \\
& Noise & 80.22 & 0.05 & 5.78 & 0.68 & 0.25 & 11.77 \\
& Perspective & 51.10 & 0.04 & 6.20 & 0.63 & 0.35 & 13.10 \\
ARAS-CSD & Rotation & 72.48 & 0.06 & 6.78 & 0.63 & 0.33 & 12.91 \\
 & Flip & 51.15 & 0.04 & 6.15 & 0.63 & 0.35 & 13.09 \\
& Resized Crop & 51.41 & 0.04 & 6.20 & 0.63 & 0.35 & 13.09 \\
& Combined & 100.79 & 0.06 & 6.62 & 0.68 & 0.25 & 11.82 \\
\midrule
& Baseline & 72.23 & 0.06 & 2.40 & 0.46 & 0.60 & 18.71 \\
& Noise & 133.36 & 0.08 & 2.67 & 0.56 & 0.43 & 15.01 \\
& Perspective & 74.68 & 0.06 & 2.40 & 0.47 & 0.60 & 18.57 \\
ARAS-CUGAN & Rotation & 96.00 & 0.05 & 3.46 & 0.48 & 0.57 & 17.64 \\
 & Flip & 69.90 & 0.05 & 2.39 & 0.50 & 0.56 & 18.01 \\
& Resized Crop & 73.70 & 0.06 & 2.40 & 0.47 & 0.60 & 18.58 \\
& Combined & 150.88 & 0.09 & 3.74 & 0.59 & 0.39 & 14.63 \\
\midrule
\midrule
& Baseline & 16.85 & 0.01 & 3.57 & \cellcolor{gray!11}0.57 & \cellcolor{gray!11}0.40 & \cellcolor{gray!11}14.06 \\
& Noise & 72.71 & 0.04 & 3.56 & \cellcolor{gray!11}0.64 & \cellcolor{gray!11}0.28 & \cellcolor{gray!11}12.43 \\
& Perspective & 16.17 & 0.01 & 3.54 & \cellcolor{gray!11}0.57 & \cellcolor{gray!11}0.40 & \cellcolor{gray!11}14.06 \\
ARAS-SGAN3 & Rotation & 43.14 & 0.02 & 4.57 & \cellcolor{gray!11}0.58 & \cellcolor{gray!11}0.38 & \cellcolor{gray!11}13.81 \\
 & Flip & 16.94 & 0.01 & 3.59 & \cellcolor{gray!11}0.57 & \cellcolor{gray!11}0.40 & \cellcolor{gray!11}14.06 \\
& Resized Crop & 16.62 & 0.01 & 3.55 & \cellcolor{gray!11}0.57 & \cellcolor{gray!11}0.40 & \cellcolor{gray!11}14.06 \\
& Combined & 92.21 & 0.05 & 4.61 & \cellcolor{gray!11}0.64 & \cellcolor{gray!11}0.28 & \cellcolor{gray!11}12.47 \\
\midrule
& Baseline & 16.68 & 0.01 & 3.70 & \cellcolor{gray!11}0.57 & \cellcolor{gray!11}0.41 & \cellcolor{gray!11}14.11 \\
& Noise & 70.68 & 0.04 & 3.69 & \cellcolor{gray!11}0.64 & \cellcolor{gray!11}0.29 & \cellcolor{gray!11}12.45 \\
& Perspective & 15.93 & 0.01 & 3.68 & \cellcolor{gray!11}0.57 & \cellcolor{gray!11}0.42 & \cellcolor{gray!11}14.11 \\
ARAS-SGAN3-D & Rotation & 42.85 & 0.02 & 4.66 & \cellcolor{gray!11}0.58 & \cellcolor{gray!11}0.39 & \cellcolor{gray!11}13.85 \\
 & Flip & 16.71 & 0.01 & 3.73 & \cellcolor{gray!11}0.57 & \cellcolor{gray!11}0.41 & \cellcolor{gray!11}14.10 \\
& Resized Crop & 16.44 & 0.01 & 3.69 & \cellcolor{gray!11}0.57 & \cellcolor{gray!11}0.41 & \cellcolor{gray!11}14.11 \\
& Combined & 91.43 & 0.05 & 4.69 & \cellcolor{gray!11}0.64 & \cellcolor{gray!11}0.29 & \cellcolor{gray!11}12.48 \\
\end{tabular}
\end{table}
\subsection{Downstream Utility}

We compared downstream utility as semantic segmentation performance (F1 score), human perceived quality score (HPQS) between 1-5 Likert scale and image quality metric (FID) in Table~\ref{tab:f1_summary_results}. F1 scores are averaged over six semantic segmentation models (UNet~\cite{ronneberger2015u}, UNet++~\cite{zhou2018unet++}, PAN~\cite{li2018pyramid}, DeepLabV3+~\cite{chen2018encoder}, SegFormer~\cite{xie2021segformer} and FPN~\cite{lin2017feature}). This comparison shows that FID does not reliably track downstream utility (F1 score). For example, datasets ARAS-train and ARAS-test show 2.1 FID score, compared to the ARAS-train augmented with CUGAN (conditionally generated UNetGAN) having 21.3 FID score meanwhile downstream task performances are 76.5 and 77.6 respectively, highlighting image fidelity, measured in FID score, is not a clear indicator of downstream utility. Similarly, human perception evaluates BELDE-K with 3.42 and BELDE-CA-NV with 3.09 mean realism scores, while BELDE-K achieves 58.3 and BELDE-CA-NV 66.2 F1 score, demonstrating that human perception is not always indicative of downstream task performance too. Moreover, when evaluated as a single dataset, CUGAN achieves the worst FID and human perception scores, but when used together with real data, CUGAN improves over the baseline performance, showing that synthetic data utility is not a singular concept to measure in isolation. 
\begin{table}[ht]
\centering
\small
\caption{Comparison of image quality (FID), Human-Perceived Quality Score (HPQS), and downstream utility (F1 Score) across different training and test configurations.}
\label{tab:f1_summary_results}
\begin{tabular}{llrcr}
\textbf{Train} & \textbf{Test} & \textbf{~~FID $\downarrow$} & \textbf{~~HPQS $\uparrow$} &\textbf{~~F1 (\%) $\uparrow$} \\
\midrule
ARAS                       & ARAS          & 2.09  & 3.54 & $76.5 \pm 0.4$ \\
SGAN3                      & ARAS          & 16.85 & 3.43 & $73.2 \pm 0.4$ \\
SGAN3-D                    & ARAS          & 16.68 & 3.28 & $74.8 \pm 0.2$ \\
CUGAN                      & ARAS          & 72.23 & 2.96 & $54.3\pm 0.7$ \\
ARAS, SGAN3, CUGAN  ~~     & ARAS          & 13.19 & 3.31 & $77.2 \pm 0.4$ \\
ARAS, CUGAN                & ARAS          & 21.27 & 3.25 & $77.6 \pm 0.7$ \\
ARAS, SGAN3                & ARAS          & 12.12 & 3.49 & $77.7 \pm 0.3$ \\
BELDE                      & BELDE         & 0.36  & 3.32 & $83.0 \pm 0.5$ \\
BELDE                      & BELDE-CA-NV~~~& 20.4  & 3.09 & $66.2 \pm 0.9$ \\
BELDE                      & BELDE-K       & 41.6  & 3.42 & $58.3 \pm 0.2$ \\
\bottomrule
\end{tabular}
\end{table}
\section{Discussion}
The empirical findings reveal a complex, often paradoxical relationship between automatic image quality metrics and human perception, challenging the validity of using standard distribution distances for domain-specific evaluation. While metrics such as FID, KID, LPIPS, SSIM, and PSNR respond consistently to severe image degradations, they do not necessarily capture the granular features that humans use when judging realism. Consequently, improvements in automated metric values should not be interpreted as direct evidence of superior perceptual quality, nor should poorer metric values be assumed to imply lower realism.

The perturbation experiments summarized in Table~\ref{tab:metrics_evaluation_reduced_target} provide evidence that most metrics are highly sensitive to substantial image distortions. Across all datasets, additive Gaussian noise and the combined perturbation consistently produced the largest degradation in automated scores. For example, ARAS-Test exhibits an increase in FID from 2.09 under the baseline setting to 54.98 under noise and 89.86 under the combined transformation. Similar trends are observed across all dataset configurations, indicating rotation, noise and combination of perturbations affect FID scoring the most, compared to flip, resized crop and perspective transformations. 

Conversely, the magnitude of metric changes fails to align with human sensitivity under safe geometric transformations. As shown in Table~\ref{tab:image_transforms}, human participants were highly robust to geometric changes, maintaining accuracies above 90\% for rotation, perspective, flip, and resized-crop transformations. Yet, these same transformations trigger inflated FID scores or volatile shifts and we hypothesize this is due to the directional sensitivity of ImageNet filters to orientation. For instance, image rotation increased the FID of ARAS-Test from 2.09 to 36.51, despite only a modest reduction in participant accuracy from 97.13\% to 94.05\%. This discrepancy suggests that common distribution-based metrics are hyper-sensitive to low-level structural changes that have negligible impact on human judgments of image realism.

Our realism assessment in Table~\ref{tab:realism_likert} further highlights this perceptual disconnect, revealing an intriguing subversion of expectations. Among all datasets, the real ARAS dataset achieved the highest realism score ($\mu=3.54$), closely followed by ARAS-SGAN3 ($\mu=3.43$) and BELDE-K ($\mu=3.42$). Notably, the synthetic ARAS-SGAN3 imagery was perceived as nearly as realistic as the real ARAS imagery, and it actively outpaced multiple real datasets such as BELDE ($\mu=3.32$) and BELDE-CA-NV ($\mu=3.09$), despite exhibiting a substantially worse FID score.

This implies that generative models have successfully learned to synthesize hyper-realistic topological features that satisfy human expectations of clear structural definition, even when real sensor data is plagued by atmospheric attenuation, blur, and lighting anomalies. This divergence is mirrored in the semantic mask verification analysis, where human accuracy is remarkably close between the real (72.87\%) and synthetic (72.41\%) domains, and average reaction times remain highly comparable (9.17~s vs. 9.28~s).

Reaction times and participant accuracy in Table~\ref{tab:image_transforms} provide additional insight into human decision-making. Participants required more time to evaluate heavily degraded images, with the combined perturbation producing the longest average response time (8.10~s) and the lowest classification accuracy (56.63\%). In contrast, the baseline condition yielded both the highest accuracy (97.13\%) and lower response times (6.35~s). This increase in cognitive latency and quickest reaction time for noise perturbation (6.04~s) suggest that uncertainty climbs as semantic coherence deteriorates and humans rely on high-level semantic and structural cues rather than low-level pixel statistics when assessing realism.

Crucially, extending this evaluation to operational value reveals that downstream task utility, quantified via semantic segmentation performance (F1 score) in Table~\ref{tab:f1_summary_results}, is heavily decoupled from both automated metrics and human perception. While a low FID is often assumed to predict better utility, augmenting the real baseline with CUGAN data degrades the FID from 2.09 to 21.27 yet paradoxically improves downstream segmentation performance from $76.5 \pm 0.4$ to $77.6 \pm 0.7$. Conversely, human preferences also fail to reliably predict operational success; participants rated BELDE-K as more realistic than BELDE-CA-NV (HPQS of 3.42 vs. 3.09), yet the segmentation model achieved a substantially higher F1 score on the less visually appealing BELDE-CA-NV ($66.2 \pm 0.9$) compared to BELDE-K ($58.3 \pm 0.2$). Furthermore, while standalone CUGAN imagery yields the poorest isolated metrics across all dimensions, its cooperative use alongside real data successfully beats the baseline. Because these misalignments run in different directions (metrics overpenalizing useful data, humans overrating less useful data), neither automatic statistics nor perceptual preference can stand in for direct measurement of downstream utility. Dataset size is also another driving factor for FID, as ARAS-train to ARAS-test measures 2.09 while BELDE-train to BELDE-test yields 0.36 FID score, indicating this automatic evaluation metric is susceptible to volume as well as fidelity as BELDE is $11\times$ the size of ARAS dataset while the compared datasets are essentially Sentinel-2 imagery.

Taken together, these findings demonstrate that optimization paths tracking purely toward FID minimization risk producing generative models that prioritize visual cleanliness at the cost of genuine data utility. Distribution-based metrics remain useful for quantifying statistical similarity between large datasets, but they should not be interpreted as direct proxies for perceptual realism or operational performance. Because downstream task utility is often decoupled from automated metric performance, evaluating generative models for Earth observation requires multiple complementary perspectives. A reliable framework should jointly consider quantitative metrics, human perception, and downstream task performance, since no single axis fully characterizes synthetic data quality.

\section{Conclusion}
\label{sec:conclusion}

In this work, we presented a study of human perception and automatic evaluation of synthetic data quality for Earth observation, revealing systematic misalignment between perceived quality, quantitative metrics, and downstream task utility. Using Earth observation datasets from multiple regions, generative models spanning Denoising Diffusion Probabilistic Models (DDPMs) and Generative Adversarial Networks (GANs), and a land-cover semantic segmentation task,  we evaluate automatic, perceptual and task performance scores under a unified framework. Through extensive testing, we find that existing automatic evaluation metrics are sensitive to different types of perturbations, exhibiting substantial variation even under semantically small augmentations. Furthermore, these metrics are not well aligned with human perception, revealing failure modes and potential vulnerability to manipulation via image perturbations. The human study also shows that real and synthetic image–mask pairs are equally interpretable to evaluators, and that some synthetic datasets are perceived as more realistic than real datasets affected by domain shift. In our evaluation setup, the StyleGAN3-based datasets were rated more realistic than the diffusion-based datasets.

Most importantly, the segmentation benchmark reveals that downstream utility is decoupled from both metrics and human perception: synthetic data that scores poorly in isolation can nonetheless improve a real-data baseline when used in combination, and lower-FID data does not reliably translate into higher segmentation performance. Optimizing generative models purely for automatic metrics therefore does not guarantee downstream utility.

These findings indicate that no single evaluation axis (automatic metrics, human perception, or task performance) is sufficient on its own, particularly under the domain shift between Earth observation imagery and the ImageNet-pretrained features underlying most automatic metrics. A promising direction for future work is the development of an evaluation metric that is robust to the perturbations destabilizing current metrics, aligned with human perception, and predictive of downstream task performance.

\section*{Acknowledgements}
Ümit Mert Çağlar is a beneficiary of the TUBITAK BIDEB 2211-Domestic Graduate Scholarship Program. The experiments reported in this work were fully performed at TUBITAK ULAKBIM, High Performance and Grid Computing Center (TRUBA).

\bibliographystyle{splncs04}
\bibliography{main}

\end{document}